| B [T] | 5 | 4.8 | 4.6 | 4.4 | 4.2 |
|---|---|---|---|---|---|
| z | 3.96 | 4.59 | 4.22 | 3.97 | 4.70 |
| ν | 0.94 | 0.81 | 0.90 | 1.04 | 0.80 |

# Vortex-Glass Melting Transition and Scaling Behavior in the Cuprate Superconductor $Nd_{2-x}Ce_xCuO_y$ at Low Temperatures

O. M. Stoll[1,2], A. Wehner[2], R. P. Huebener[2], M. Naito[3]

[1]*Present Address: University of California San Diego, Physics Department 0319, 9500 Gilman Dr., La Jolla, CA 92093, U.S.A.*

[2]*Physikalisches Institut, Lehrstuhl für Experimentalphysik II, Universität Tübingen, Morgenstelle 14, D- 72076 Tübingen, Germany*

[3]*NTT Basic Research Laboratories, 3-1 Morinosato-Wakamiya, Atsugi-shi, Kanagawa, 243-01, Japan*

Corresponding author:   O. M. Stoll
                        Address: University of California San Diego, see above
                        E-mail: ostoll@physics.ucsd.edu
                        Phone: 858-534-7161
                        Fax: 858-534-0173

# Abstract

We have investigated the vortex-glass transition for the cuprate superconductor Nd$_{2-x}$Ce$_x$CuO$_y$ (NCCO) in the region of low temperatures (T < 4.4 K) and moderate up to high magnetic fields (3 T < B < 5 T) compared to the upper critical field B$_{c2}$ (B$_{c2}$ ≈ 6 T for NCCO). We analyzed our data using the vortex-glass theory and found an excellent scaling behavior of the current-voltage characteristics in this temperature region. The critical exponents have been calculated and were compared to the few available investigations for NCCO and to the results for other materials. Additionally we determined the melting line for low temperatures and compared it to the available literature data. In this way we could draw a complete picture of the mixed-state phase diagram for this material. The knowledge about the phase diagram is crucial for the understanding of the transport properties of the high-T$_c$ cuprates in the mixed state. It is found that for NCCO, thermal fluctuations play an important role even at temperatures below 4.2 K. This is possibly due to the relatively high anisotropy of this material. An extrapolation procedure is proposed in order to overcome our experimental limitation of the temperature range and to obtain the melting line for T > 4.4 K. A kink in the extrapolated portion of the melting line is found and explained in terms of a loss of universality of the critical exponents for temperatures T > 6 K.

# 1. Introduction

In the conventional (low-Tc) superconductors, the phase diagram is well described by the Ginzburg Landau theory. In magnetic fields between the lower critical value B$_{c1}$ and the upper critical value B$_{c2}$ flux-lines penetrate the sample and form the Abrikosov vortex lattice. In the cuprate superconductors thermal fluctuations, the short coherence lengths and the strong anisotropy lead to modifications of this phase diagram. It can show a rich diversity of lattice, glass or liquid phases that dramatically influence the electrical transport behavior of the superconductors. The knowledge about the details of the phase diagram is therefore extremely interesting both for possible applications of the high-T$_c$ superconductors and for fundamental science.

Since the prediction of the vortex-glass (VG) by Fisher *et al.* [1] there has been a strong experimental effort to investigate the liquid-to-glass transition and the glass phase itself in various high-Tc superconductors (HTS). Most of the measurements have been done for $YBa_2Cu_3O_{7-\delta}$ (YBCO) (see e.g. [2,3,4]) but also other cuprate materials like $Bi_2Sr_2Ca_2Cu_3O_{10}$ (BiSCCO) [5,6] and $Tl_2Ba_2CaCu_2O_8$ [7], conventional superconductors like Al [8], $Mo_3Si$ [9] or the cubic superconductor $(K,Ba)BiO_3$ [10] have been investigated. Surprisingly, for NCCO there exist only a few measurements conducted exclusively in the high temperature, low magnetic field region [11,12].

For a long time NCCO was thought to represent an anomaly within the family of the copper oxide superconductors. Convincing measurements of the magnetic penetration depth (see e.g. [13]) and tunnel spectroscopy (see e.g. [14]) gave evidence for NCCO to be an electron doped s-wave single gap BCS-superconductor. In contrast to this, recent phase sensitive experiments [15] point to d-wave symmetry of the superconducting order parameter for this material. These results show that many aspects of the electron doped cuprate superconductors remain still unclear. NCCO has a relatively low $T_c$ of about 24 K, which should reduce the influence of thermal fluctuations. On the other hand its larger mass anisotropy ($\gamma^2 = m_c^*/m_{ab}^* \approx 600$) compared to YBCO (factor 30) and $La_{2-x}Sr_xCuO_y$ (LSCO, factor 170) should enhance vibrations of the vortex lines around their equilibrium positions. It is believed that the weak interplane coupling for NCCO originates from the missing apical oxygen atoms in this material. Due to the tetragonal crystal structure of NCCO, there are no twin boundaries present. Therefore the overall pinning force is expected to be relatively weak compared to other cuprates. Oxygen vacancies [16] and the Ce cations which are randomly distributed over the Nd sites [11] have been discussed as possible sources for intrinsic disorder in this material. Because of its small $B_{c2}$ of around 6 T, the whole phase diagram of NCCO is experimentally accessible, in sharp contrast to most of the HTS, which usually have $B_{c2}$ values of 100 T or more. It appears that NCCO is an ideal prototype material for the study of vortex-matter and its different phases. Especially measurements at low temperatures can give clues about the relative importance of thermal fluctuations and anisotropy on the various phases of the vortex-matter for the cuprates.

In this paper we report on electrical resistance measurements for NCCO at low temperatures (T < 4.4 K) and moderate up to high magnetic fields B ≈ $B_{c2}$ (3 T to 5 T). We show that a vortex-glass to -liquid transition can be observed in NCCO at temperatures below 4.4 K. The critical exponents and the glass line for this temperature region are determined and compared to the high temperature data available in literature. In this way the phase diagram for this material could be extended to lower temperatures.

## 2. **Theoretical Expressions**

According to the conventional theory of vortex motion at small driving forces (small current densities j), the Anderson-Kim model, there should be an ohmic resistance at all nonzero temperatures due to thermal activation of flux-lines out of their pinning potential wells. This linear resistivity $\rho_l$ vanishes exponentially with decreasing temperature following the relation $\rho_l \propto \exp[-U/k_B T]$ (U is the activation energy and $k_B$ is Boltzman's constant) [17]. According to this a type-II superconductor in a magnetic field larger than $B_{c1}$ is never truly superconducting in the thermodynamic sense, except for T = 0. It has an exponentially small but finite linear resistivity $\lim_{j \to 0} E/j = \rho_l$. The major flaw of the Anderson-Kim theory is, that it assumes an uncorrelated thermally activated motion of vortex-bundles of a finite size $V_c$ disturbing the phase coherence of the macroscopic wavefunction and therefore causing electric resistivity.

The vortex glas model introduced by Fisher *et. al.* proposes that there exists a truly superconducting vortex-glass phase below a critical temperature $T_g$ in which the linear resistivity $\rho_l$ vanishes in the sense $\lim_{j \to 0} E/j = 0$. The transition between the liquid and the glass is assumed to be of second order. The fundamental assumption of this model is that the interaction between vortex lines cannot be neglected and persists over a distance $\xi_{vg}$ which can by far exceed the Larkin-lengths $L_c$ and $R_c$ of the collective pinning theory, defining the size of the independently hopping vortex bundles [18,19]. The characteristic length scale over which the phase coherence of the macroscopic pairing wave function is retained is expected to diverge at the phase boundary $T = T_g$:

$$\xi_{vg} \propto |T - T_g|^{-\nu} \qquad (1)$$

ν is the static critical exponent. Additionally, near second order phase transitions a critical slowing of the dynamics is expected:

$$\tau_{vg} \propto \xi_{vg}^z \propto |T - T_g|^{-zv} \tag{2}$$

z is the dynamical critical exponent. Roughly speaking, $\tau_{vg}$ is the time it takes a fluctuation of the size $\xi_{vg}$ to relax. The critical exponents are universal numbers and should therefore depend only on the nature of the phase transition (or more precisely on its universality class) and not on the details of the material.

One possible experimental access to the different phases of the vortex-matter is provided by electrical resistance measurements. The glass theory predicts a characteristic behavior of the current voltage characteristic (IVC) which can be divided in three temperature regions [20]:

(1) $T = T_g$: Exactly at the glass temperature Fisher *et al.* predict a ‚critical' power law behavior:

$$E \propto j^{(z+1)/(D-1)} \tag{3}$$

D is the dimension of the system.

(2) $T > T_g$ (vortex-liquid phase): For small current densities $j < j_{nl}$ a nonvanishing linear resistivity is expected:

$$E \propto \rho_l j \tag{4}$$

For $j > j_{nl}$ the critical behavior (3) is recovered.

(3) $T < T_g$ (vortex-glass phase): The vortex-glass phase is characterized by diverging pinning barriers $U(j) \propto U_0(j_c/j)^\mu$ ($j_c$ is the critical current density, $\mu < 1$ is the glassy exponent) that hamper the motion of vortices for $j \to 0$ so that a vanishing linear resistivity can be expected. For small current densities $j < j_{nl}$, this leads to:

$$E \propto \exp[-(j_c / j)^\mu] \qquad (5)$$

For $j > j_{nl}$ one finds again the critical behavior (3).

Another important consequence of continuous phase transitions is a common functional dependence of the relevant physical properties at temperatures close to the transition. According to this, the IVCs should collapse onto two universal functions when being scaled with the appropriate powers of $\xi_{vg}$. For a three dimensional system, this can be expressed in the scaling relation:

$$\rho \cdot \xi_{vg}^{z-1} \approx \Re_\pm (j \cdot \xi_{vg}^2) / j \cdot \xi_{vg}^2 \qquad (6)$$

$\Re_\pm$ are the universal functions for temperatures above (+) and below (-) the glass temperature respectively.

From the requirement that all isothermal IVCs for $T > T_g$ (4) should fall on one common curve and by inspection of equation (6) one can easily see that the temperature dependences of $\rho_l$ and $j_{nl}$ must be given by the relations

$$\rho_l \propto \xi_{vg}^{-(z-1)} \propto (T - T_g)^{\nu(z-1)} \qquad (7)$$

and

$$j_{nl} \propto 1/\xi_{vg}^2 \propto (T - T_g)^{2\nu}. \qquad (8)$$

For temperatures close to but still above $T_g$, the vortex-glass correlation length $\xi_{vg}$ is very large and dominates all other physical length scales. In this case the transport behavior of the system should be well described by the vortex-glass model. One can define a length $\xi_d$ over which the applied current density j can affect thermal distributions:

$$j\xi_d^2\phi_0 \approx k_B T \qquad (9)$$

$\phi_0$ is the magnetic flux quantum. At small current densities vortices are excited over dimensions that are much larger than $\xi_{vg}$. The correlation is disturbed which leads to the appearance of linear resistivity $\rho_l$. If the length $\xi_d$ is much smaller than $\xi_{vg}$, the critical behavior (3) dominates [21]. For $T \to T_g$ and the condition $\xi_{vg} = \xi_d$ equation (9) leads directly to the expression (8) for the crossover current density $j_{nl}$.

The lower critical dimension for the existence of the vortex-glass state lies between 2 and 3 [20]. For two-dimensional samples, no freezing into a truly superconducting glass state is expected. In this case the linear resistivity shows the typical ‚Anderson-Kim' behavior: $\rho_l \propto \exp[-(T_0/T)]$. It is interesting that the vortex-glass model offers no possibility for the determination of the glass line $B_g(T)$ itself. For this purpose, different models like the simple ‚cage model' [22,23] which neglects the effect of flux pinning, have to be used.

## 3. **Experimental Procedures and Setup**

Our samples were high-quality epitaxial c-axis oriented thin films of NCCO grown by molecular beam epitaxy on $SrTiO_3$ substrates [24, 25]. The film thickness was 90-100 nm. In the following we present results obtained for two different samples. Sample 1 was a slightly overdoped NCCO film ($x \approx 0.16$) with a critical temperature $T_c = 21.3$ K. The resistivity values at 295 K and 30 K were $\rho(295\ K) = 167.2\ \mu\Omega cm$ and $\rho(30\ K) = 22.7\ \mu\Omega cm$, respectively. Sample 2 was optimally doped ($x \approx 0.15$) with $T_c = 23.9$ K and $\rho(295\ K) = 140.0\ \mu\Omega cm$, $\rho(30\ K) = 21.0\ \mu\Omega cm$. The width of the superconducting transition $\Delta T_c$ was less than 1 K in both cases. The IVCs were measured using a four-point microbridge geometry fabricated by a standard photolithographic process. The effective dimensions of the microbridges were $w = 30\ \mu m$, $L = 100\ \mu m$ (w = width, L = distance between the voltage leads) for sample 1 and $w = 30\ \mu m$, $L = 300\ \mu m$ for sample 2. Ag contact pads have been deposited on the NCCO films and two 25 $\mu m$ diameter Al wires were bonded to each pad. As confirmed in separate experiments, the contact resistance was less than 50 mΩ per Al wire (length ≈ 3 mm) and independent of the applied current up to 100 mA per wire.

In our measurements we swept the current at slow rates of about 0.5 mA/s using a low noise battery powered current source controlled by an external ramp generator. The sample voltage was amplified with a model SRS 560 preamplifier and recorded with a personal computer at a sampling rate of 50 Hz. The slow sweeping rates allowed us to use a 30 Hz lowpass filter for the voltage and for the current signal in order to obtain time averaged IVCs with a low overall noise level. Therefore a sufficient voltage signal could be obtained even in the low measuring current region.

During all measurements the samples were immersed in liquid helium to provide the most effective cooling in order to minimize any influence due to Joule heating. The drawback of this method is the restriction of the temperature range to values below 4.2 K. For the isothermal measurements the temperature was always lowered between the recording of two subsequent IVCs, beginning with 4.2 K, in order to assure a thermal equilibrium in the helium bath. The temperature was adjusted by pumping and its value was obtained from the helium vapor pressure. However, in some measurements temperatures up to 4.4 K have been realized by slightly raising the pressure in the cryostat. In these cases a reasonably long equilibration time was allowed before the actual measurements have been started.

## 4. Experimental Results

### 4.1 Scaling Behavior of the IVCs

Fig. 1 shows a typical series of IVCs at various temperatures and B = 4.8 T. The resistivity $\rho = E/j$ is plotted double logarithmically versus the measuring current I. In this way one can identify immediately the ohmic behavior at temperatures $T > T_g$ and the vanishing critical current, which are the fingerprints of the liquid phase. The curvature of the IVCs for temperatures below $T_g$ is negative, as predicted by the vortex-glass theory. Here we have a truly superconducting phase with a nonzero critical current $j_c \neq 0$ and $\lim_{j \to 0} E/j = 0$. The temperature where the behavior of the IVCs changes (dashed line in Fig. 1) defines the vortex-glass transition. Despite the low temperature values, one can clearly see the phase transition. For this sample, the described behavior could be experimentally observed within the magnetic field range 3.9 – 5 T. For smaller magnetic fields, the phase transition lies beyond our highest accessible temperature of about 4.4 K.

The data was quantitatively analyzed according to Fisher's scaling theory. One important prerequisite for the application of this theory is that the system can be considered as 3D. In 2-dimensional systems no critical scaling is expected. We will return to this issue in the following section and present arguments showing that it is indeed reasonable to consider the NCCO films as 3D. In order to scale the data, the critical exponents have to be determined first. We obtained the dynamical critical exponent z from the slope of the IVCs at the phase transition (T = $T_g$) in the double logarithmic plot using equation (3) for the case of a 3-dimensional system (D = 3). The statical critical exponent ν can be determined using the temperature dependence (7) of the linear resistivity $\rho_l$ for T > $T_g$. Plotted double logarithmically versus (T-$T_g$) this should yield a straight line with the slope ν(z-1). Fig. 2 shows such a plot for B = 5 T. Table 1 summarizes the results of the values for the critical exponents for sample 2. From the inset of Fig. 2 we see that there is no systematic variation of the critical exponents with the magnetic field. The average values are $\bar{z} = (4.29 \pm 0.34)$ and $\bar{\nu} = (0.90 \pm 0.10)$. From this we can also calculate the slope $\bar{\nu}(\bar{z}-1) = (2.96 \pm 0.45)$. Taking these values, the IVCs could be scaled by applying equation (6) for the 3-dimensional case and converting the resistivity and current axis using

$$(V/I)_{sc} = (V/I) \mid T - T_g \mid^{-\nu(z-1)} \text{ and } I_{sc} = I \cdot \mid T - T_g \mid^{-2\nu} \qquad (10)$$

The result is shown in Fig. 3. For the moment we concentrate on the black curves. The figure contains 25 IVCs covering a temperature range between 1.9 K and 4.4 K at three different magnetic fields (4.2, 4.4 and 4.8 T). As expected from the scaling theory, the IVCs collapse clearly on two distinct branches. The scaling behavior proves the universal nature of the critical exponents. To test the stability of the scaling we repeated this procedure with the same data, this time using values for the critical exponents which are purposely set off the determined averages by the amount of one error bar. The results using z = (4.29 + 2·0.34) = 4.97 and ν = (0.90 +

2·0.10) = 1.10 are shown in the inset of Fig. 3. We clearly see that the quality of the scaling deteriorates in this case. This shows that the obtained critical exponents provide a fairly accurate result. The value of the parameter $T_g$, which had to be determined in order to scale the IVCs, yields the liquid-solid transition line in the phase diagram for NCCO and is discussed in the following section.

The gray curves in Fig. 3 represent magnetic fields between 1 and 3.8 T for which the phase transition lies beyond the experimentally accessible temperature range. These curves are based on the same critical exponents as the other data, this time using $T_g$ as a free fitting parameter. With this extrapolation procedure the value of $T_g$ could be determined even for temperatures beyond 4.4 K, assuming that the critical exponents do not change. The results for $T_g$ and the melting line are discussed in the following section.

In order to prove the consistency of the data, the statical critical exponent $\nu$ was determined independently using equation (8). According to (8) the current density $j_{nl}$, defining the point in the IVC where nonlinearity sets in, should vary with the inverse square of the correlation length $\xi_{vg}$. Plotting $j_{nl}$ double logarithmically versus the normalized temperature $(T-T_g)$ should therefore yield a straight line with the slope $2\nu$. The result of this analysis is shown in Fig. 4. Here we have chosen $I_{nl}$ as the current at which the nonlinear resistivity has increased 20% beyond its ohmic value:

$$\rho(I_{nl}) = 1.2\rho_l \tag{11}$$

The major drawback of this method is a relatively high inaccuracy in reading the datapoints, as reflected in the large error bars. Nevertheless, a linear fit to the average values and the determination of the statical exponent yields $\nu \approx 1$, which is in good agreement with the previously mentioned values (see Tab. 1 and Fig 2). It appears so far that the vortex-glass theory gives consistent results.

Because of the universal nature of the critical exponents it is interesting to compare them to the few results that have been obtained for the material NCCO up to now and also to the data for other materials. According to the theory, the critical exponents depend only on the universality class of the phase transition and not on the specific details of the superconductor. The theoretical estimate for the 3D vortex-glass transition is $z = 4 – 6$ and $\nu \approx 1\text{-}2$ [20].

For Bose-glasses, slightly higher dynamical exponents $z_{BG} = 1/2 \cdot (3 \cdot z + 1) = 6 - 10$, $\nu_{BG} > 1$ [26,27] can be expected. For conventional spin-glasses, which are believed to be in the same universality class as the vortex glasses, the statical exponent is somewhat smaller $z_{SG} = 4$, $\nu_{SG} \approx 1/2$ [20]. To our knowledge, measurements on NCCO have been exclusively conducted at high temperatures (10 – 24 K) and low magnetic fields (B < 1 T) [11,12]. For the dynamical exponent Yeh *et al.* find a small value of $z = 3$ [11]. On the other hand Roberts *et al.* report a higher value of $z = 5.4$ [12]. It appears that our value for the dynamical exponent lies in between these two references. For the statical exponent both references find larger values of $\nu = 2.1$ [11] and $\nu = 1.75$ [12] compared to our analysis. In YBCO the range spreads from $z = 4.3 – 4.8$ for the dynamical and from $\nu = 1 – 2$ for the statical exponent in good agreement with the theoretical work and with numerical simulations (see e. g. [2, 3, 4]). Recent experiments with untwinned proton irradiated YBCO single crystals find $\nu(z-1) = 5.1 \pm 0.5$ [28] which is a bit larger than our numbers. Interestingly, in this work a threshold for point disorder has been proposed, below which a first order transition rather than a second order transition can be found. For our intrinsically untwinned NCCO films, the observation of a second order transition could mean that the amount of point disorder is high enough to exceed this threshold. Another interesting material to compare is the $(K,Ba)BiO_3$ cubic superconductor [10]. This isotropic system has a similar $T_c$ of around 30 K. Here exponents of $z = 5$ and $\nu = 1$ are found. The dynamical exponent and the value $\nu(z-1) = 3.9$ are both slightly larger than in our experiments. We conclude that our data is well comparable to the results for YBCO and in agreement with the theoretical work. Recent experiments on proton irradiated YBCO and on $(K,Ba)BiO_3$ find somewhat larger values of the dynamical critical exponents. Compared to the previous measurements on NCCO we find lower values of the statical exponent and a dynamical exponent that lies in the middle of the available references [11] and [12].

## 4.2 Phase Diagram for NCCO

In order to gain insight into the relative importance of thermal fluctuations and anisotropy on the vortex-matter in NCCO, one has to investigate the B-T phase diagram for this material. Therefore in Fig. 5 we determined the $B_g(T_g)$ diagram using the glass temperatures $T_g$ obtained for different magnetic fields in the previous scaling analysis. The data shown is normalized with the upper critical field $B_{c2}(0)$ and the critical temperature $T_c$. For $B_{c2}(0)$ we used the linear extrapolation of the melting line to T = 0 yielding $B_{c2}(0)$ = 5.92 T for the slightly overdoped film (sample 1, solid squares in Fig. 5) and $B_{c2}(0)$ = 6.37 T for the optimum doped film (sample 2, open squares in Fig. 5). Note that these values are in good agreement with the data found in ref. [29] for optimum and overdoped NCCO-films. Fig. 5 includes also the extrapolated melting temperatures $T_g$ for sample 2 at lower magnetic fields (crossed squares in Fig. 5). From Fig. 5 we see that the data for the two different samples show good quantitative agreement for T < 4.2 K. The melting line can be described by the ‚cage-model' [23, 22] which assumes that the vortexline is moving in a harmonic potential defined by the surrounding nearest neighbor lines. In general the $B_g(T)$ curve follows a power law of the form

$$B_g \propto (1-T/T_c)^m \qquad (12)$$

with m ≈ 2 in the simplest case (T ≈ $T_c$) [30]. The double logarithmic plot in Fig. 6 shows that in the low temperature region, such a power law with m = (2.19 ± 0.11) is a reasonable approximation. We note that there is no discrepancy between the actually measured and the extrapolated data at the point of intersection ($T_{cr}/T_c$ ≈ 0.2). However, we see a distinct kink (arrow in Fig. 6) in the melting line at the reduced temperature $T_{cr}/T_c$ ≈ 0.25 ($T_{cr}$ ≈ 6 K and $B_{cr}$ ≈ 3 T). Here the exponent m changes from m ≈ 2.19 to m = (0.95 ± 0.05). The reason leading to this kink is discussed later in this section.

The data for NCCO available in literature is almost exclusively limited to temperatures well above 10 K. In order to get a complete picture of the B-T phase diagram for NCCO and to compare this work with the other results, we added some of the literature data to Fig. 5. Fabrega *et al.* find an irreversibility line with m = 2 on NCCO single crystals [16]. Extrapolating their fitting function to reduced temperatures $T/T_c$ < 0.25, we find a good agreement with our measurements both for the qualitative and quantitative behavior. However, Hwang *et al.* report a somewhat lower power of m = 1.7 [31] measured at single crystals and Yeh *et al.* find a still lower value of m = 1.2 for thin films [11].

In the following we discuss the implications of our results for the general understanding of vortex-matter in NCCO at low temperatures. First of all we note that the data of Yeh *et al.*, which show an almost linear behavior, at first sight seem to be consistent with our data obtained with the extrapolation procedure mentioned in section 4.1. This could lead to the assumption, that the kink observed in the melting line is a property specific for thin NCCO-films. On the other hand, the thin film data of Roberts *et al.* [12] seem to coincide with the single crystal measurements in [16] which rules out this possibility. Often a change in the functional dependence of the $B_g(T)$ lines is assumed to be related to a sudden change of the dimensionality of the system, e.g. in the strong coupled superconductor YBCO, relatively small values of m = 1.3 – 1.5 are observed [4] whereas 2D melting in strongly anisotropic systems like BiSCCO can lead to much larger values for m [5, 30]. As anticipated, no change in the slope of the $B_g(T)$ line has been found in the isotropic system $(K,Ba)BiO_3$ [10]. To estimate the magnetic field $B_{2D}$ at which a Kosterlitz-Thouless transition can be expected we use the expression [30]

$$B_{2D} \approx \frac{\phi_0}{\delta^2 \gamma^2} \qquad (13)$$

Here, $\delta$ is the distance between two adjacent copper oxide planes and $\gamma$ is the anisotropy factor. Using $\delta$ = 6.04 Å [32] and $\gamma^2$ = 600 for NCCO we obtain $B_{2D} \approx$ 9.2 T. Recalling that the upper critical field $B_{c2}$ is about 6 T, we see that no decoupling transition can be expected within the entire mixed state of NCCO. Finally we have to compare the correlation length $\xi_{vg}$ of the vortex-glass theory to the other relevant length scales. As long as $\xi_{vg}$ is smaller than the film thickness d we deal with a 3D system. If $\xi_{vg}$ becomes smaller than the intervortex distance $a_0$ the prerequisite of the glass theory, namely an interaction between the vortices over distances larger than $a_0$, is void. In this case the critical exponents loose their universality and are expected to change with the magnetic field. This has been observed in different material systems [2, 4, 8, 12]. The inset of Fig. 6 shows an estimate of the correlation length $\xi_{vg}$ obtained by taking the experimental values for $I_{nl}$ and using equation (9) for two different magnetic fields. The solid line is a fit with the expected temperature dependence $\xi_{vg}(T) = A\,|1-T/T_g|^{-\nu}$ according to equation (7). For $T_g$ and $\nu$ we used the previously determined values $T_g \approx 2.1$ K (B = 4.5 – 5 T) and $\nu \approx 1$ while A was left as the only fitting parameter. In this way we extrapolated the data to higher and lower Temperatures. For A we obtain A = (38.2 ± 2.0) nm. We see that in this first approximation $\xi_{vg}$ stays smaller than the film thickness down to the lowest measured temperature, which is consistent with the observation of a 3D scaling behavior for the IVCs. For the intersection with the $a_0$-line for the case $B_{cr} = 3$ T ($a_0 \approx 25$ nm) we find approximately T = 5.7 K which is in good agreement with the corresponding temperature $T_{cr} = 6$ K at which the kink in the melting line appears. From this we conclude that the kink is due to a loss of universality of the critical exponents, which begin to vary with the magnetic field. For our extrapolation procedure we implicitly assumed the critical exponents to be universal numbers. Therefore we cannot apply this procedure at temperatures higher than approximately 6 K. This also explains the surprisingly good match of the actually measured data points with the extrapolated ones at the point of intersection. At this temperature there is no doubt about the universal nature of the critical exponents.

To complete the picture of the B-T phase diagram for NCCO in Fig. 5 we added the $B_{c2}(T)$-line from Ref. [33]. We see that there is a narrow region occupied by the vortex-liquid state, which extends to low temperatures. Thermal fluctuations seem to play an important role in this cuprate superconductor although there are no indications for a decoupling of the $CuO_2$-planes. This is consistent with the results for (K,Ba)BiO$_3$. However in this material the origin of the large vortex liquid phase at low temperatures remains unresolved [10]. In NCCO the relatively large anisotropy factor clearly promotes the thermal vibrations of the vortex-lines. In this respect, NCCO occupies an intermediate position between the strongly anisotropic materials like the Tl-superconductors or BiSCCO and the less anisotropic materials like YBCO and LSCO.

## 5. Summary and Conclusion

The vortex-glass transition and the phase diagram for the cuprate superconductor NCCO have been studied at temperatures below 4.4 K and at magnetic fields between 3.9 and 5 T. For this purpose we measured the IVCs for this material with the sample immersed in liquid helium. Using slow sweep rates and low pass filters together with a high gain substantially reduced background noise. We showed that even at low temperatures a glass-to-liquid transition can be found. The data was analyzed using the vortex-glass theory and the critical exponents have been determined. We found values of $\bar{z} = (4.29 \pm 0.34)$, $\bar{\nu} = (0.90 \pm 0.10)$ and $\bar{\nu}(\bar{z}-1) = (2.96 \pm 0.45)$ in good agreement with the theoretical predictions and with the results for YBCO. The few available literature data for NCCO reports higher values for $\nu$, whereas the critical exponent seems to lie in the middle of the above mentioned references. Scaling with these exponents showed that the IVCs collapse nicely on two distinct branches. The consistency of the scaling analysis was also checked using the temperature dependence of $I_{nl}$ and by varying the obtained critical exponents.

The $T_g$ values previously determined for the scaling analysis were used to plot the low-temperature phase diagram for NCCO. With a new extrapolation procedure we were able to determine the $T_g$ even for magnetic fields where the melting point lies beyond the experimentally accessible temperature range. The phase diagram is in nice agreement with the high temperature data found in literature except for a kink in the melting line at $T_{cr} \approx 6$ K and $B_{cr} \approx 3$ T. The kink could be explained by a loss of universality of the critical exponents which was implicitly assumed for the extrapolation process. We found no indications for a Kosterlitz-Thouless transition in the mixed state of NCCO. It seems that in this material thermal fluctuations of the flux lines around their equilibrium positions are of high relative importance probably due to the relatively high anisotropy when compared to YBCO or LSCO. However, many properties of this material, like the symmetry of the order parameter are still under discussion and are obviously requiring more experimental and theoretical work.

## **Acknowledgements**

This work was financially supported by the Deutsche Forschungsgemeinschaft. O. M. S. acknowledges financial support from the Deutscher Akademischer Austauschdienst (DAAD).

## Figure Captions:

Fig 1: Double logarithmic plot of the resistivity vs. measuring current I for temperatures between 4.4 and 1.9 K. The dashed line indicates the vortex-glass to liquid transition. The temperature values are: 4.4, 4.2, 4.1, 3.9, 3.8, 3.4, 3.1, 2.6, 2.3 and 1.9 K.

Fig. 2: Plot of $\log(\rho_l)$ vs. $\log[(T - T_g)/T_g]$ for the determination of the critical exponent $\nu$ for the magnetic field B = 5 T. A linear fit yields the slope $\nu(z - 1) = 2.796$. The inset shows the exponents $\nu$ and $z$ plotted vs. magnetic field.

Fig. 3: Plot of the scaled resistance $(V/I)_{sc}$ vs. scaled current $I_{sc}$ (see eq. (10)) for 3 different magnetic fields and for temperatures between 1.9 and 4.4 K (black curves). The gray curves represent smaller magnetic fields (1 – 3.8 T) and were obtained by the extrapolation procedure explained in the text. Here $T_g$ was used as a fitting parameter. The figure contains a total of 34 curves. The inset shows the same data scaled with exponents which are calculated by adding twice the standard deviation to the average values yielding $z = 4.97$, $\nu = 1.1$.

Fig. 4: Double logarithmic plot of the average $I_{nl}$ defined by equation (11) vs. normalized temperature $[(T - T_g)/T_g]$ for 2 magnetic fields. The dashed lines are linear fits to the average values.

Fig. 5: Phase diagram for NCCO. The square symbols represent the results of this work. The solid line is the irreversibility line of ref. [16] extrapolated to lower temperatures. The triangular symbols are for single crystalline samples and the circles for thin film samples. The crosses are the results for the Bc2(T)-line taken from ref. [33]. The different phases of the vortex-matter are indicated.

Fig. 6: Double logarithmic plot of the melting line. Two regions with different exponents m = (2.19 ± 0.11) and m = (0.95 ± 0.05) are clearly visible. The arrow indicates the transition between these regions. In the inset, the correlation length $\xi_{vg}$ is plotted vs. temperature. The dotted lines

indicate the film thickness d and the intervortex distance $a_0$ for $B_{cr}$ = 3 T. The solid line is a fit with the function $\xi_{vg} = A \cdot |1-(T/2.1\ K)|^{-1}$ (2.1 K is the glass-temperature for B = 3 T).

Tab. 1: Values for the statical and the dynamical exponent for sample 2 in the magnetic field range 4.2 – 5 T.